\documentclass[11pt]{article}

\usepackage{amsmath, amssymb}
\usepackage{graphicx}
\usepackage{natbib}
\usepackage{authblk}   
\title{Drift, diffusion and divergence}

\author{Laurette S. Tuckerman}
\affil{
    Physique et M\'ecanique des Milieux H\'et\'erog\`enes (PMMH), CNRS, ESPCI Paris, PSL University, Sorbonne Universit\'e, Universit\'e de Paris, 75005 Paris, France\\ \texttt{laurette@pmmh.espci.fr}}

\begin{document}

\maketitle

\begin{abstract}
  Turbulent Taylor-Couette flow displays traces of axisymmetric Taylor
  vortices even at high Reynolds numbers. With this motivation,
  \cite{Feldmann} carry out long-time numerical simulations of
  axisymmetric high-Reynolds-number Taylor-Couette flow.  They find
  that the Taylor vortices, using the only degree of freedom that
  remains available to them, carry out Brownian motion in the axial
  direction, with a diffusion constant that diverges as the number of
  rolls is reduced below a critical value.
\end{abstract}

\section{Introduction}
In \citeyear{Taylor_1923}, \citeauthor{Taylor_1923}
published his ground-breaking experiment and
linear stability calculation, whose agreement demonstrated the
validity of the Navier-Stokes equations. Since then, Taylor-Couette
flow has served as one of the protypical systems in fluid dynamics.
In the Taylor-Couette experiment, fluid is confined between two
concentric cylinders which rotate at different angular velocities. 
In laminar Taylor-Couette flow, the motion is purely azimuthal
and fluid particles at different radii do not mix.
Increasing the angular velocity difference past a critical value
leads to the formation of Taylor vortices, toroidal rolls
in which circular motion in the meridional $(r,z)$ plane
redistribute fluid and angular momentum between the radii.

Ever since Taylor described and explained the onset of
axisymmetric Taylor-vortex flow, an extravagant profusion 
of three-dimensional patterns of extraordinary variety, beauty, and
complexity have been discovered experimentally and numerically
\citep[e.g.,][]{Andereck,Chossat_Iooss,weisshaar1991twist,altmeyer2012symmetry,deguchi2013fully,akinaga2018tertiary}.
The mathematics of what is called variously 
equivariant bifurcation theory, symmetry, and pattern formation
has been brought to bear to predict and explain these 
spirals and ribbons, twists and waves, modulation and bursts.

Turbulence in Taylor-Couette flow has also been studied, 
both at high Reynolds number
and in the transitional range at low Reynolds number
\citep[e.g.,][]{Coles,Prigent_PRL,Mutabazi,Shi,lemoult2016directed}. 
But who would have thought that there was something new to be learned
about turbulence from {\it axisymmetric} Taylor-Couette flow?

\section{Summary of Paper}

It has long been known that the Taylor-vortex structure persists even
far into the turbulent regime, i.e. that turbulence is superposed on
Taylor vortices \citep[e.g.,][]{lathrop1992transition,dong2007direct,huisman2014multiple,grossmann2016high}; long-time averaging accentuates the features of these
ghostly vortices. \cite{eckhardt2020exact} have argued that under
certain hypotheses, transport of angular momentum by chaotic
fluctuations in axisymmetric Taylor-Couette flow reproduces the
transport associated with the axisymmetric component of turbulent
solutions to the full three-dimensional equations.  This suggests that
the axisymmetric problem could be viewed, not merely as a first step
towards turbulence (laminar $\rightarrow$ axisymmetric Taylor-vortex
flow $\rightarrow$ three-dimensional patterns $\rightarrow$
turbulence), but as a model for its mean (necessarily axisymmetric)
properties.  \cite{Feldmann} have carried out long-time axisymmetric
simulations of Taylor-Couette flow as a possible route towards
studying turbulent structures.

\begin{figure}
\centering
\includegraphics[width=\columnwidth]{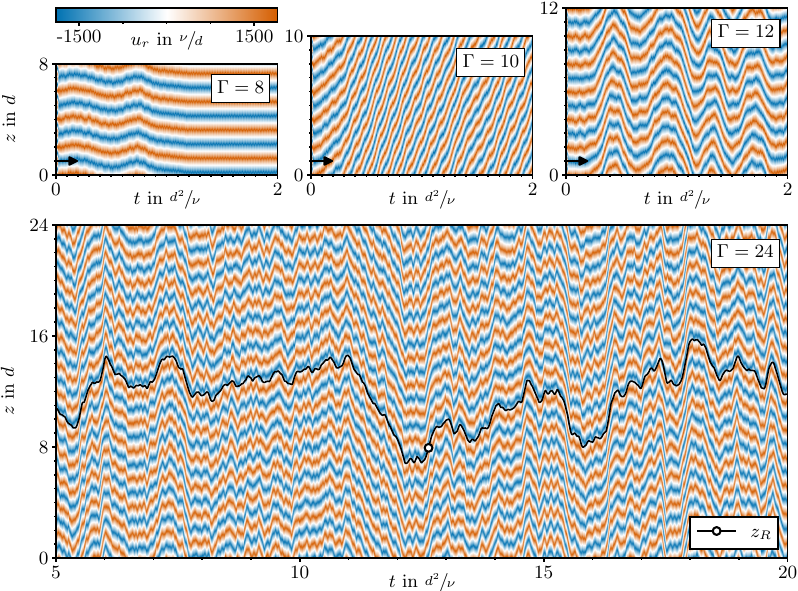}
\caption{Temporal evolution of radial velocity along an axial line at mid-gap.
  The aspect ratio $\Gamma$ of axial length to radial gap corresponds to the number 
  of vortices. For $\Gamma=8$, after an initial transient, the vortices do
  not move, while for $\Gamma=10$, they move very quickly in one direction.
  For $\Gamma=12$ and 24, the vortices sporadically change their direction
  of motion. From \cite{Feldmann}.}
\label{fig:Avila2}
\end{figure}

Axisymmetric Taylor-vortex flow consists of an axial stack of toroidal
vortices. The vortices are approximately circular, so that the number of
vortices is close to the axial-length-to-radial-gap ratio $\Gamma$.
\cite{Feldmann} observe that the number of vortices remains
constant over the course of a simulation. Such a one-dimensional
periodic structure is highly constrained and so its possible dynamics
are limited: the only remaining possible motion is an axial jiggle 
or drift of the entire stack of vortices.
\cite{Feldmann} find that for a relatively long system the rolls carry
out diffusive drift (Brownian motion) so that the variance of the
phase grows linearly in time.  Moreover, the effective diffusion
coefficient diverges following a power law as a threshold axial length
(or number of rolls) $\Gamma_c$ is approached from above.
For a shorter axial length, although there may be an immediate
adjustment of the position,
the rolls quickly becomes quasi-stationary,
with only weak chaotic motion about a fixed location.
For the parameters used by \cite{Feldmann}, $\Gamma_c=10$;
see figure \ref{fig:Avila2}.
The significance of this sharp threshold is unknown. 

Although this is an interesting puzzle by itself, its importance is
increased by its generality.  Many hydrodynamic systems are driven by
an imposed gradient of some quantity.  Rolls appear as a means of
redistributing this quantity: azimuthal or streamwise velocity for
Taylor-Couette, plane Couette or Poiseuille flow, temperature for
Rayleigh-B\'enard convection, concentration for a binary fluid.  Drift
has been observed in these other systems
\citep{xi2006azimuthal,kreilos2014comoving} and according to
\cite{Feldmann}, the drift appears to be of the same type.

Exploiting the analogy between axisymmetric Taylor-Couette flow and
two-dimensional Rayleigh-B\'enard convection \citep{veronis1970analogy},
\cite{eckhardt2020exact}
have proposed a mapping from the two Reynolds numbers (inner and 
outer, or equivalently, shear $Re_S$ and rotation $R_\Omega$)
of Taylor-Couette flow \citep{Dubrulle_2005} to the single Rayleigh number
$Ra$ of Rayleigh-B\'enard convection.
\cite{Feldmann} have provided support for this analogy
by showing that the diffusion coefficient of the axial drift was the same for
different parameter pairs $(Re_S,R_\Omega)$ yielding the same value of $Ra$.

This demonstrates the interest in axisymmetric Taylor-Couette flow from a
scientific point of view. However, the imposition of axisymmetry also
has the great advantage of economy.  Measuring diffusion coefficients
of the axial drift requires extremely long times, especially if other
parameters are varied as well, i.e.\ the number of rolls and the
Reynolds numbers.  \cite{Feldmann} have been able to measure these diffusion
coefficients because axisymmetric simulations require only a small fraction of the
time that would be required to simulate the three-dimensional flow.

One might associate axial drift (motion of the phase) with axial flux
(motion of fluid particles). To investigate this, \cite{Feldmann} have
compared simulations in which the axial flux is set to zero with those
in which the net axial pressure gradient is zero. Either condition is
valid for a periodic direction, but the choice has significant
consequences if the flow is not reflection symmetric 
\citep[e.g.,][]{edwards1991periodic}. \cite{Feldmann} find that in the absence
of axial flux, the drift is considerably reduced, but still undergoes
Brownian motion.

\section{The Future}
Several questions are raised by this paper. The most obvious and
perplexing is the reason for the abrupt threshold.
Why are shorter columns tranquil and why are slightly longer columns
suddenly so jittery? What physical phenomenon could be responsible for
such a sharp distinction?

The second question concerns its generality. \cite{Feldmann} have
given convincing evidence that the rolls in other flows, such as
Poiseuille flow, Rayleigh-B\'enard convection and Taylor-Couette flow with
no axial flux, also undergo diffusive drift. Does drift in these flows
also have a length threshold?  Are the threshold and the power law
decay exponent the same?

The third question concerns the applicability of these axisymmetric
results to the three-dimensional turbulence which naturally occurs at
these high values of Reynolds or Rayleigh number. \cite{eckhardt2020exact}
suggest that some global properties of three-dimensional turbulent
Taylor-Couette flow could be captured by its axisymmetric analogue.
Is axial drift one of those properties? What other properties might obey this?



\end{document}